\documentclass[fleqn,usenatbib]{mnras}

\usepackage{newtxtext,newtxmath}
\usepackage[T1]{fontenc}

\DeclareRobustCommand{\VAN}[3]{#2}
\let\VANthebibliography\thebibliography
\def\thebibliography{\DeclareRobustCommand{\VAN}[3]{##3}\VANthebibliography}

\usepackage{graphicx}	% Including figure files
\usepackage{amsmath}	% Advanced maths commands
\usepackage{threeparttable}

\let\oldAA\AA
\renewcommand{\AA}{\text{\normalfont\oldAA}}
\title[IC/CMB and high-z blazars]{The Impact of the CMB on the Evolution of high-$z$ Blazars}

\author[Ighina et al.]{
L. Ighina$^{1,2}$\thanks{E-mail: lighina@uninsubria.it},
A. Caccianiga$^1$,
A. Moretti$^1$,
S. Belladitta$^{1,2}$,
R. Della Ceca$^1$,
and A. Diana$^{3}$
\\
$^{1}$INAF, Osservatorio Astronomico di Brera, via Brera 28, 20121 Milano, Italy\\
$^{2}$DiSAT, Universit\`a degli Studi dell' Insubria, via Valleggio 11, 22100, Como, Italy\\
$^{3}$Dipartimento di Fisica G. Occhialini, Universit\`a degli Studi di Milano-Bicocca, Piazza della Scienza 3, 20126 Milano, Italy\\
}

\date{Accepted XXX. Received YYY; in original form ZZZ}

\pubyear{2021}

\begin{document}
\label{firstpage}
\pagerange{\pageref{firstpage}--\pageref{lastpage}}
\maketitle

\begin{abstract}
Different works have recently found an increase of the average X-ray-to-radio luminosity ratio with redshift in the blazar population.
We evaluate here whether the inverse Compton interaction between the relativistic electrons within the jet and the photons of the Cosmic Microwave Background (IC/CMB) can explain this trend.
Moreover, we test whether the IC/CMB model can also be at the origin of the different space density evolutions found in X-ray and radio selected blazar samples. By considering the best statistically complete samples of blazars selected in the radio or in the X-ray band and covering a large range of redshift (0.5$\lesssim$ \textit{z}$\lesssim$ 5.5), we evaluate the expected impact of the CMB on the observed X-ray emission on each sample and then we compare these predictions with the observations. 
We find that this model can satisfactorily explain both the observed trend of the X-ray-to-radio luminosity ratios with redshift and the different cosmological evolutions derived from the radio and X-ray band. Finally, we discuss how currently on-going X-ray missions, like eROSITA, could help to further constrain the observed evolution at even higher redshifts (up to \textit{z}$\sim$6-7).
\end{abstract}


%%%%%%%%%%%%%%%%%%%%%%%%%%%%%%%%%%%%%%%%%%%%%%%%%%

%%%%%%%%%%%%%%%%% BODY OF PAPER %%%%%%%%%%%%%%%%%%

\section{Introduction}

Blazars are radio-loud (RL\footnote{We consider a source to be radio loud if it has a radio loudness R\textgreater10, where R is defined as the ratio between the 5~GHz and 4400~\AA ~rest-frame flux densities, R=$S_\mathrm{{5 GHz}}/S_\mathrm{{4400\AA}}$ \citep{Kellerman1989}.}) Active Galactic Nuclei (AGNs) with a relativistic jet directed close to our line of sight.
Thanks to this particular orientation, the relativistic amplification of the jetted radiation makes blazars extremely bright sources and thus visible even in the primordial Universe, with the most distant currently observed at $z$=6.1 \citep{Belladitta2020}.
Interestingly, from the observation of one of these sources we can infer the total number of RL AGNs at the same redshift with similar properties, but with a misaligned jet: N$_\mathrm{{tot}}\approx$ N$_\mathrm{{obs}}\times2\Gamma^2$ (e.g. \citealt{Sbarrato2015}), with $\Gamma$ the bulk Lorentz factor of the plasma within the jet ($\Gamma\sim$5-15, e.g. \citealt{Ghisellini2015b}).
Combined, these two features make blazars ideal tools to study the cosmological evolution of Supermassive Black Holes (SMBHs) hosted in RL AGNs and to understand the role that relativistic jets play in their growth.\\
Studies of X-ray selected samples of blazars (2\textless $z$\textless 4) found that the space density of the most X-ray luminous flat spectrum radio quasars (FSRQs, hereafter simply blazars\footnote{In this work we only consider the FSRQ and not the BL Lac class of blazars. Indeed only a few BL Lac sources have been observed at high redshift (e.g. \citealt{Paliya2020b}).}) has a unique cosmological evolution, with a peak at larger redshift (\textit{z}$\sim$3.5-4, \citealt{Ajello2009, Toda2020}) compared to the radio-quiet (RQ) population \citep{Hopkins2007,Shen2020}, even considering the X-ray brightest RQ AGNs (\textit{z}$\sim$2.5, \citealt{Aird2015}). 
This result led different authors (\citealt{Ghisellini2013,Sbarrato2015}) to suggest that the formation and the growth of the most massive (M$_\mathrm{{BH}}$\textgreater$10^9$M$_{\odot}$) RL AGNs is faster compared the RQ population with similar masses, implying a strong connection between the jet and the accretion process. However, analogous studies performed on radio selected samples of blazars found that their density evolution in the radio band is similar to the one observed in the RQ population (\citealt{Mao2017,Caccianiga2019}), even when considering the most massive AGNs (M$_\mathrm{{BH}}$\textgreater$10^9$M$_{\odot}$, Diana et al. in prep.). 
Since in the spectral energy distribution (SED) of blazars both the X-ray and the radio bands are dominated by the emission produced within the relativistic jet, such a discrepancy is not obvious and its understanding is of utmost importance if we want to reliably use blazars as tracers of the SMBHs evolution.\\
One possible explanation could be that the typical X-ray-to-radio ratio (X/R) increases with redshift. Indeed such an evolution has been observed at high redshift, firstly in very radio-powerful AGNs \citep{Wu2013} and then also in the specific class of blazars \citep{Ighina2019}, where it was found that $z$\textgreater4 blazars have on average X/R ratios $\sim2$ times larger than low-$z$ ones (\textit{z}$\sim1$). One possibility often invoked in the recent literature that predicts this type of trend is that is that the increase of the X/R ratios is due to an enhanced X-ray emission caused by the interaction between the relativistic electrons in the jet and the photons of the Cosmic Microwave Background (CMB). 
In this case we expect a fraction of the electrons in the extended parts (a few kpc, e.g. \citealt{Paliya2020}) of the jet to interact via inverse Compton scattering with the CMB photons (IC/CMB) at any given redshift. Since the CMB energy density grows as U$_\mathrm{{CMB}}\propto(1+z)^4$, this process is expected to become more important at higher redshifts and thus to increase the observed X/R ratios.
Moreover, this type of X-ray luminosity evolution with redshift would also increase the number of observed blazars in flux limited X-ray surveys (and thus their space density) in a similar way to what was found by \cite{Ajello2009}. 
Even though other interpretations have been proposed to explain the observed discrepancies (see e.g., \citealt{ Wu2013}), the IC/CMB model has acquired more and more popularity after the launch of the \textit{Chandra} X-ray observatory and the discovery of many extended X-ray jets, even at high redshift (e.g. \citealt{Tavecchio2000,Siemiginowska2007,Schwartz2020,Napier2020}). Nevertheless, a scenario where all the X-ray and $\gamma$-ray extended emission is solely related to an IC/CMB process has been challenged several times by observations in the local Universe (e.g. \citealt{Georg2006,Meyer2014,Breiding2017}), with only two sources having met the IC/CMB predictions \citep{Meyer2019}. 
Interestingly, both these objects are blazars, suggesting that orientation plays an important role in the observability of the IC/CMB emission (e.g. \citealt{Simionescu2016}).
In any case, we still expect the IC/CMB interaction to take place even at scales not resolved by current X-ray facilities, especially at high-$z$. The real question is the relative amount of X-ray radiation produced trough this process %, %{\bf also at scales unresolved by current X-ray facilities},
with respect to the total one observed. For this reason, a \textit{fractional} IC/CMB model has been recently proposed \citep{Wu2013}, where the interaction with the CMB is responsible for only a small fraction ($\sim$3\%) of the total X-ray emission  at low redshift (\textit{z}$\sim$1, e.g. \citealt{Zhu2019}) becoming more and more important at $z$\textgreater3-4,  where it can be dominant.\\
In this work, by using a statistical approach, we want to verify whether the X-ray enhancement observed in high-z blazars can be explained by a \textit{fractional} IC/CMB model and if this same type of evolution can also explain the differences observed in the space density of blazars obtained from X-ray and radio surveys.

The paper is structured as follows: in Sec. \ref{sec:samples} we present the blazar samples and relative data used throughout the work; in Sec. \ref{sec:IC/CMBevo} we discuss the expected X/R ratios evolution considering an IC/CMB scenario and how it compares to observations; in Sec. \ref{sec:density} we derive the expected X-ray space density evolution from the IC/CMB model and compare it with observations; in Sec. \ref{sec:predictions} we show how this model can be tested with currently on-going X-ray missions; finally, in Sec. \ref{sec:conclusions} we summarise our results and conclusions.\\
We assume a flat $\Lambda$CDM cosmology with $H_{0}$=70 km s$^{-1}$ Mpc$^{-1}$, $\Omega_m$=0.3 and $\Omega_{\Lambda}$=0.7. Spectral indices are given assuming $S_{\nu}\propto \nu^{-\alpha}$ and all errors are reported at 1$\sigma$, unless otherwise specified.

%%%%%%%%%%%%%%%%%%%%%%%%%%%%%%%%%%%%%%%%%%%%%%%%%%%%%%%%%%%%%%%%%%%%%%%%%%%%%%%%%%%%%%%%%%%%%%%%%%%%%%%%%%%%%%%%%%%%%%%%%%%%%
\section{Blazar Samples}
\label{sec:samples}
In order to study the evolution of blazars across cosmic times in this work we considered three different blazar samples which, together, cover an extended range of redshifts, from the local Universe, \textit{z}$\sim$0.5, up to \textit{z}$\sim5.5$. A full description of the first two samples (C19 and BZ1.5Jy) is given in \cite{Caccianiga2019} and \cite{Ighina2019} respectively, here we report only a short summary of them.
\begin{figure}
\centering
\includegraphics[width=\linewidth]{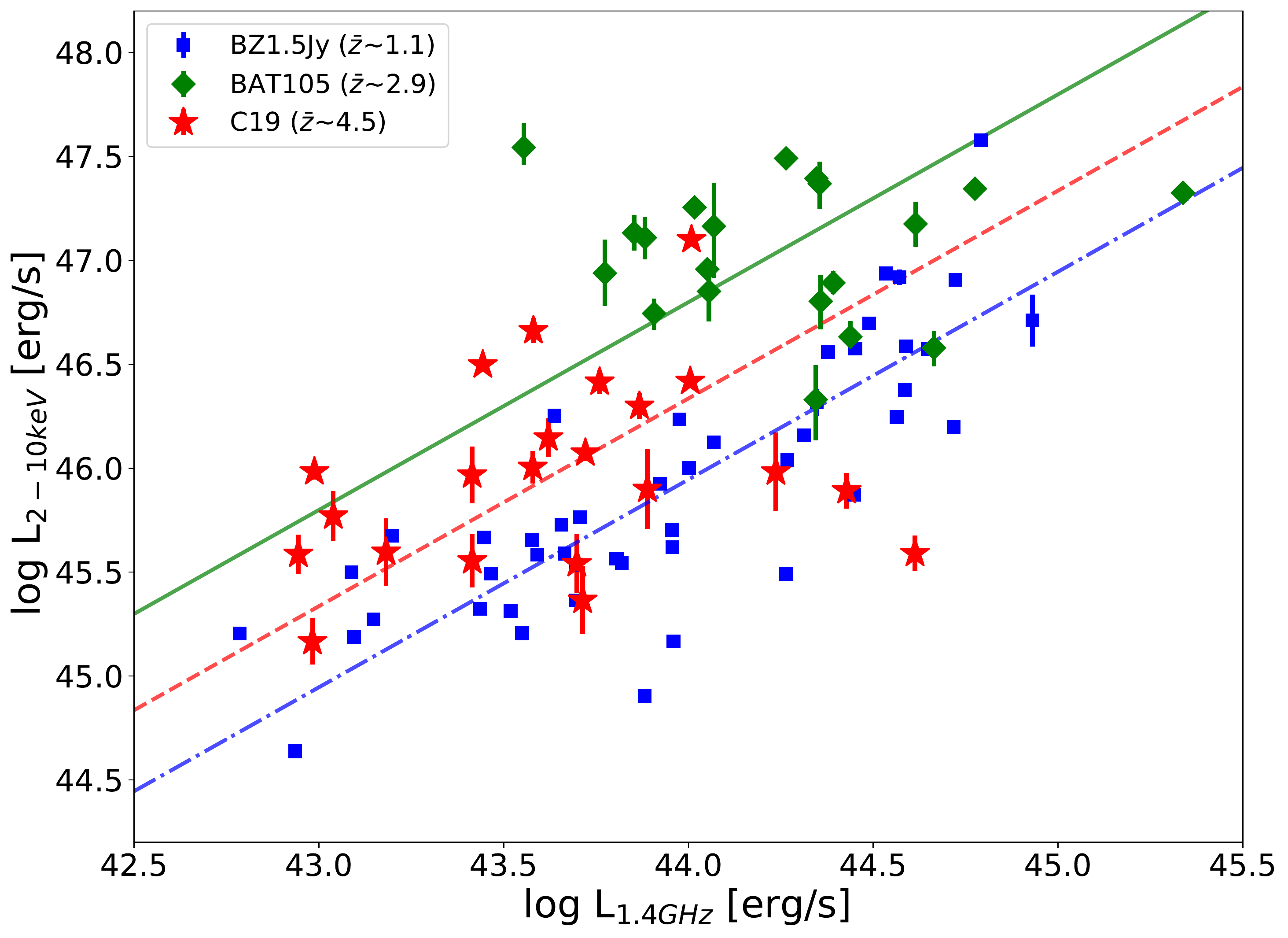}
\caption{X-ray luminosity in the 2-10 keV energy band as a function of the radio luminosity at 1.4 GHz. The sources are represented with different symbols according to the sample they belong to: blue squares = BZ1.5Jy, red stars = C19 and green diamonds = BAT105. Beside the name of each sample in the legend, we also report the mean redshift of each sample in brackets. The errors on the X-ray luminosities are at 90\% confidence level. The three lines represent the mean X-ray-to-radio luminosity ratio of the given sample: dashed-dotted blue for BZ1.5Jy, \textless log(L$_\mathrm{X}$/L$_\mathrm{R}$)\textgreater=1.95; dashed red for the C19, \textless log(L$_\mathrm{X}$/L$_\mathrm{R}$)\textgreater=2.34; solid green for BAT105, \textless log(L$_\mathrm{X}$/L$_\mathrm{R}$)\textgreater=2.80.}
\label{fig:lum_range}
\end{figure}

\begin{itemize}
    \item \textbf{C19 sample}: it is the largest flux-limited (S$_{5GHz}$\textgreater30 mJy) sample of blazars at \textit{z}\textgreater4. The sources have been selected from the Cosmic Lens All Sky Survey (CLASS; \citealt{myers2003}) in the radio band and then followed-up spectroscopically \citep{Caccianiga2019} in order to have a confirmation of their high-$z$ nature. After completing the X-ray coverage of the sample through dedicated observations we analysed their X-ray properties and identified the bona-fide blazars accordingly \citep{Ighina2019}\footnote{We note that the last blazar (GB6J1711+3830) was observed and classified as such after the publication of \cite{Ighina2019}, therefore we report in Appendix \ref{app:GB6J1711+38} the analysis of its \textit{Swift}-XRT observation.}. The final sample is composed by 22  objects (over a total of 25 candidates) with mean redshift \textit{z}$\sim$4.5.\\
    
    \item \textbf{BZ1.5Jy sample}: in order to have a reference sample at lower redshift and in the same range of radio luminosities of the C19 one, we considered all the blazars (FSRQs) in the 5$^{th}$ edition of the BZCAT catalogue \citep{Massaro2015} with a radio flux density S$_{1.4GHz}$\textgreater 1.5 Jy.
    At these high flux density levels almost all blazars have been discovered and identified and thus this can be confidently considered as a radio flux-limited sample with radio luminosities similar to the C19 sample. Moreover, with such an high flux limit the large majority of these objects has already been observed (and detected) in the X-rays, thus avoiding any possible bias against X-ray weak blazars. Finally, we considered the ones observed by the \textit{Swift}-XRT telescope, which have radio properties similar to the rest of the sources, and analysed their X-ray spectra. There are a total of 46 sources in this sample, with mean redshift \textit{z}$\sim$1.1;\\
    
    \item \textbf{BAT105 sample}: The sources of this sample were selected from \textit{The 105-Month \textit{Swift}-BAT All-sky Hard X-Ray Survey} catalogue \citep{Oh2018}, which is the collection of all the \textit{Swift}-BAT observations (in the 14-195 keV band) during its first 105 months of activity. Since we are interested in the X-ray properties of high redshift blazars, we started by considering only the sources with \textit{z}\textgreater2. We then selected all the objects with an X-ray flux $f_\mathrm{{14-195keV}}> 8.40 \times 10^{-12}$ erg s$^{-1}$ cm$^{-2}$, above which the sky coverage can be considered approximately constant (see Fig. 11 in \citealt{Oh2018}) and the catalogue complete over 90\% of the entire sky. Then, in order to select all the bona-fide blazars we considered all the objects with a radio counterpart in the NRAO VLA Sky Survey (NVSS, \citealt{condon1998}). We note that adopting this approach we are also including two sources (SWIFT~J0909.0+0358 at $z$=3.288 and SWIFT~J0131.5-1007 at $z$=3.515) that are not classified as blazars in the BAT catalogue, but which, after its publication, were identified as such \citep{Paliya2019,Marcotulli2020} according to their broad-band properties. In the following we will also consider the soft X-ray (0.3-10~keV) properties of these sources as reported in the 2$^\mathrm{nd}$ version of the \textit{Swift}-XRT Point Source catalogue (2SXPS; \citealt{Evans2020}). In total there are 20 objects in this sample, with a mean redshift of \textit{z}$\sim$2.9.

\end{itemize}

In Fig. \ref{fig:lum_range} we show the X-ray (2-10~keV) and radio (1.4~GHz) rest-frame luminosity distributions together with the mean X-ray-to-radio luminosity ratio (X/R=log$\frac{L_\mathrm{{2-10keV}}}{L_\mathrm{{1.4GHz}}}$) for each of the three samples described above. From the plot, it is clear that the three samples have on average significantly different ratios. We note that both the X-ray and radio luminosities refer to the indicated rest-frame frequencies (i.e. they have been $k$-corrected). In order to compute the rest-frame X-ray luminosities we adopted the photon index measured in the 0.3-10~keV observation of each object, which also covers the 2-10~keV rest-frame band in the redshift range here considered, $z$=0-5.5. In the radio band we used the observed spectral index for the C19 sample (0.15-1.4~GHz, \citealt{Caccianiga2019}), while we assumed a flat spectral index ($\alpha_r$=0), which is typical of FSRQs \citep{Padovani2017}, for the BZ1.5Jy and the BAT105 samples. We note that by adopting a flat radio spectrum also for the entire C19 sample, their X/R ratios distribution does not significantly change (the shift in the mean value is \textless0.05).
For these reasons, the X-ray and radio slopes can be considered relatively well constrained and, therefore, the observed discrepancy in the X/R ratios at low and high-$z$ cannot be attributed to a different redshift-dependent sampling of the rest-frame spectra. In the following we will test whether the observed evolution in the X/R values  can be explained by the interaction of the relativistic jets with the photons from the CMB.

%%%%%%%%%%%%%%%%%%%%%%%%%%%%%%%%%%%%%%%%%%%%%%%%%%%%%%%%%%%%%%%%%%%%%%%%%%%%
\section{Redshift evolution of the X-ray-to-radio luminosity ratios}
\label{sec:IC/CMBevo}

\subsection{IC/CMB expected trend}
As previously mentioned, different independent studies of the most powerful RL AGN population (i.e. mostly blazars) found a similar increase of the X-ray emission as a function of redshift when compared to both the optical emission (e.g. \citealt{Wu2013, Zhu2019}) and the radio one (e.g. \citealt{McKeough16, Ighina2019}). In this work we want to verify whether the observed evolution can be explained as due to a \textit{fractional} IC/CMB interaction by focusing on well-defined samples of blazars. According to this model, only a small fraction of the total X-ray luminosity is produced via the IC/CMB mechanism, which, however, largely depends on redshift since it follows the CMB energy density trend, U$_\mathrm{{CMB}}\propto$(1+z)$^{4}$.
At the same time we are also expecting an analogous decrease of the radio emission produced by the same electrons, which are quickly cooled down by the IC interaction with the CMB (e.g. \citealt{Schwartz2019}). However, this kind of effect can be considered negligible in the blazar class, since their observed radio emission at high frequencies is dominated by the beamed photons produced in the compact inner-most region of the jet, close to the accreting SMBH (e.g. \citealt{Ghisellini2009}), where the extremely large magnetic field energy density overpowers the CMB one even at high redshift (e.g. \citealt{Afonso2015}). Therefore, while the overall radio emission can be considered to remain unchanged with respect to redshift, the fraction of X-ray emission produced through an IC/CMB interaction will become more relevant or even dominant at high-$z$, even if it is negligible at low redshift.
According to this model, the overall X-ray-to-radio luminosity ratio of blazars can be expressed as:
\begin{equation}
\frac{L_X}{L_R}(z)\: = \: \frac{L_X}{L_R}(z=0) \times \,  \left[(1 - A_{0}) + A_0 (1 + z)^4\right]
    \label{eq:enhace}
\end{equation}
Where A$_0$ represents the fraction of the X-ray emission produced by the interaction with CMB photons at $z$=0. In the following we consider the X-ray luminosity integrated in the 2-10~keV band while the radio one at 1.4~GHz, unless otherwise specified.\\
In order to explain the observed difference between the low-$z$ and high-$z$ X/R ratios, the fraction of X-ray luminosity produced by the IC/CMB interaction at $z$=0, should be, as expected, very small on average: A$_0\approx$0.0016 \citep{Ighina2019}, but the corresponding amplification factor (1-$A_0$ +$A_0$(1+z)$^4$) would be $\sim$2 at \textit{z}$\sim$4.\\
Moreover, even though we can expect the overall population to follow on average the redshift trend predicted by the IC/CMB model, the importance of this interaction may vary from source to source depending on the physical properties of each SMBH and jet, and thus resulting in a scatter around the mean relation (e.g. \citealt{Cheung2004}).
Indeed, as shown in Fig. \ref{fig:lum_range}, a large scatter around the mean X/R values is present also in all the three samples considered here. 
For this reason, we improved the analysis of the previous work by assuming that the impact of the CMB interaction is not the same for all sources, but, instead, it follows a distribution of values which we derived from observations. As we will see below the assessment of this scatter is necessary to understand the extreme features observed in the BAT105 sample. 

%%%%%%%%%%%%%%%%%%%%%%%%%%%%%%%
\subsection{Comparison with observations}
\begin{figure*}
\centering
\includegraphics[width=\linewidth]{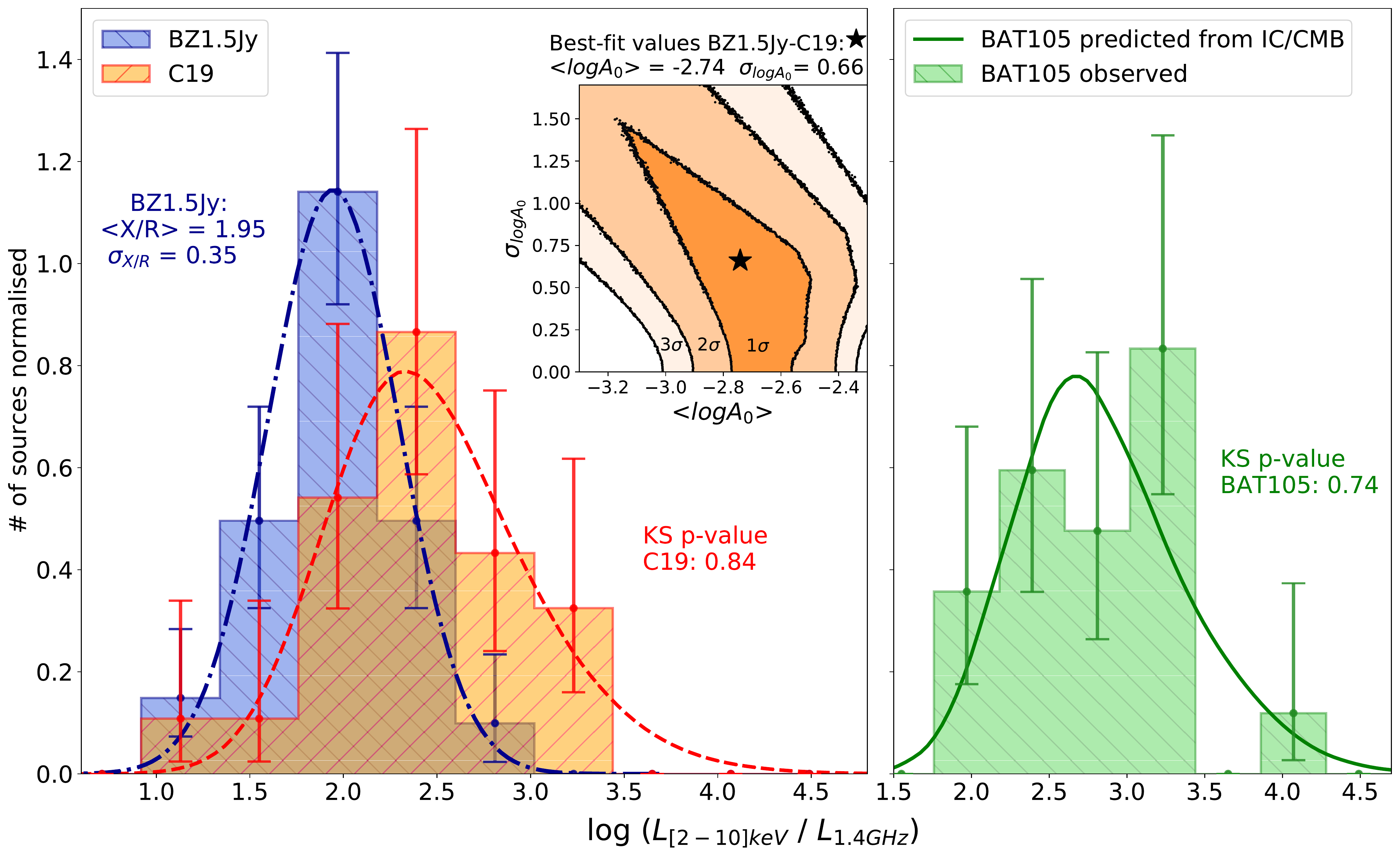}
\caption{\textbf{Left panel}: observed X/R distributions used to the derive the \textless logA$_0$\textgreater and $\sigma_\mathrm{{logA_0}}$ parameters: the BZ1.5Jy sample (filled blue histogram) with its best-fit log-normal distribution (dashed-dotted blue line) and the C19 sample (filled orange histogram) with the its best-fit X/R distribution computed from eq. \ref{eq:enhace} at $z$=4.5 (dashed red line). In the top-right corner we report the 1, 2 and 3$\sigma$ confidence contours for the \textless logA$_0$\textgreater and $\sigma_\mathrm{{logA_0}}$ parameters obtained from the comparison between the IC/CMB model and the X/R ratios distribution in the C19 sample. The best-fit parameter is shown with a black star. \textbf{Right panel}: comparison between the observed (filled green histogram) X/R distributions for the \textit{Swift}-BAT survey at 2\textless $z$\textless5 and the expected one (solid green line) considering an IC/CMB model \textit{with the parameters derived from the samples in the left panel}. We stress that the green curve is not a fit to the BAT data but it is the prediction of the model described in the text. The error-bars indicate the Poisson errors associated to each bin of the distributions.}
\label{fig:distrXR_cmb}
\end{figure*}
\begin{figure}
\centering
\includegraphics[width=0.95\linewidth]{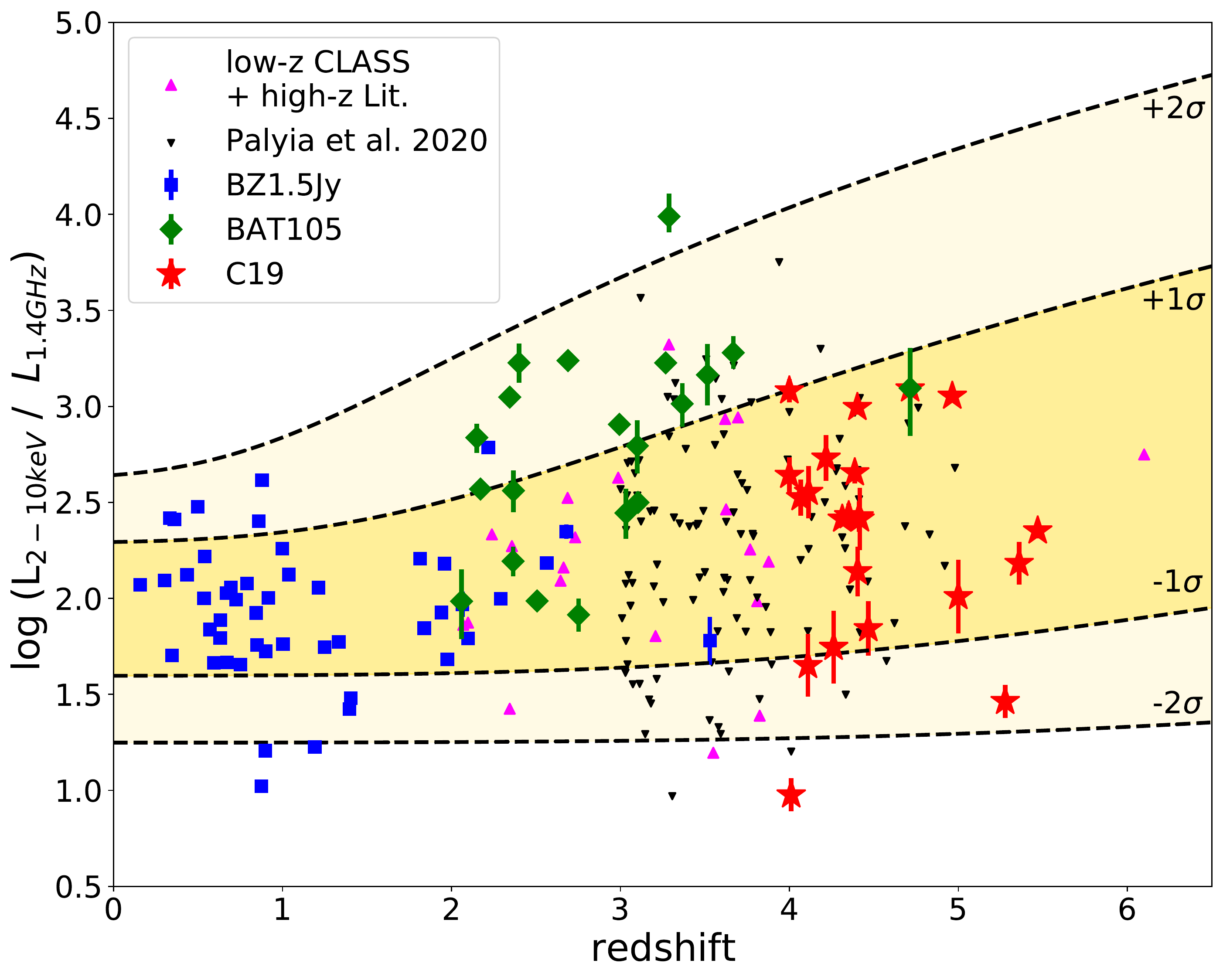}
\caption{Expected evolution of the overall X/R ratios as a function of redshift for a radio selected sample of blazars considering an initial distribution with: \textless X/R\textgreater = 1.95, $\sigma_{X/R}=0.35$ and a distribution for the logA$_0$ parameter described by: \textless logA$_0$\textgreater=-2.74, $\sigma_{logA_0}$=0.66. The dashed black lines represent the expected evolution for a source with both the initial X/R and logA$_0$ distant 1 or 2 $\sigma$ from the mean values. The different data points represent the samples the sources belong to: BZ1.5Jy (blue squares), BAT105 (green diamonds) and C19 (red stars). We also report $z$\textgreater2 blazars taken from the literature: magenta triangles are 2\textless $z$\textless 4 blazars from the CLASS survey that were serendipitously observed by the XMM-\textit{Newton} or \textit{Swift}-XRT telescopes plus the only $z$\textgreater6 confirmed blazars \citep{Belladitta2020}; black triangles are bona-fide blazars discussed in \citep{Paliya2020}.}

\label{fig:xr_ev}
\end{figure}

In order to determine the redshift evolution of the X/R ratios, we started by assuming that both the initial X/R values at $z$=0 and the A$_0$ parameter follow a log-normal distribution. The first is constrained by the observed X/R values at \textit{z}$\sim$1, in the BZ1.5Jy sample\footnote{This is valid if A$_0$ is, as expected, very small. Condition which we have checked a posteriori.} (\textless X/R\textgreater=1.95 and $\sigma_\mathrm{{X/R}}$=0.35). In order to estimate the best distribution describing the A$_0$ parameter, we computed the expected X/R values at $z$=4.5 (mean redshift of the C19 sample) from eq. \ref{eq:enhace} for different \textless logA$_0$\textgreater \, and $\sigma_\mathrm{{logA_0}}$.
Comparing the expected distributions with the values observed in the C19 sample through the Kolmogorov-Smirnov (KS; \citealt{Smirnov1948}) test we found that the best parameter set is given by: \textless logA$_0$\textgreater=-2.74 and $\sigma_\mathrm{{logA_0}}$=0.66 (KS p-value=0.84). The two theoretical X/R distributions just described are reported in Fig. \ref{fig:distrXR_cmb} (left panel) together with the ones actually observed in the BZ1.5Jy and C19 samples. 
It is interesting to note that in the case of $\sigma_\mathrm{{logA_0}}$=0 we have A$_0\approx$0.002, similar to the one estimated in \cite{Ighina2019} where we had assumed a single value of A$_0$ instead of a distribution.
In Fig. \ref{fig:xr_ev} we report the expected X/R evolution for a radio selected sample of blazars as a function of redshift considering the distribution of initial values observed in the BZ1.5Jy and the best-fit logA$_0$. The evolution of sources having an initial X/R ratio and logA$_0$ parameter both within 1$\sigma$ and 2$\sigma$ from the central values is highlighted (yellow regions and dashed black lines). We also report the X/R values for the sources in the three samples here considered as a function of their redshift, together with other confirmed blazars at \textit{z}\textgreater2 from the literature.\\
While the mean redshift of the BAT105 sample is \textit{z}$\sim$3, its X/R mean value is higher compared to the C19 sample at \textit{z}$\sim$4.5. Indeed it is clear from Fig. \ref{fig:xr_ev} that the BAT105 sample is largely composed by sources where both the initial X/R and the A$_0$ values are \textgreater 1$\sigma$ far from the mean ones. This is likely a selection bias which favours the inclusion of the extreme X-ray sources.
In order to quantitatively test this hypothesis, we started from the radio luminosity function (RLF) derived by \citealt{Mao2017} (hereafter Mao17)\footnote{Computed using a sample of radio selected blazars at 1.4~GHz in the range $z$=0.5-3. In this work we considered the RLF derived from the ``clean sample" discussed in Mao17.} by creating a mock sample where redshift and radio flux/luminosity are known for each source. We then considered the X/R distribution predicted by the IC/CMB model (using the best-fit values derived above) at the redshift of each source to estimate their X-ray luminosity. In particular, for each simulated object at redshift $z$, we randomly extracted an X/R value from the predicted distribution at that redshift and then computed the corresponding X-ray luminosity given the radio luminosity of the source. 
In order to convert the X-ray luminosity from the 2-10~keV into the 14-195~keV and to compute the corresponding observed flux, we assumed a photon index $\Gamma=1.54$. This is the mean value obtained in \cite{Ricci2017} by analysing the broad-band XRT-BAT spectra (0.3-150~keV, observed frame) of all the FSRQs in the 70 month \textit{Swift}-BAT catalogue \citep{Baumgartner2013}.
Finally, we apply to the mock sample the same X-ray flux limit of the BAT105 sample: $f_\mathrm{{14-195 keV}}$\textgreater8.40 $\times$ 10$^{-12}$ erg cm$^{-2}$ s$^{-1}$. In Fig. \ref{fig:distrXR_cmb} (right panel) we show the BAT105 sources (filled green histogram) together with the simulated sample generated as explained above (solid green line).\\
By comparing the predicted and observed distribution for X-ray selected samples with the KS test we found that an IC/CMB model, \textit{with the parameters fixed to those that best explain the X/R observed in the C19 sample}, naturally predicts the presence of objects with extreme X-ray properties already at \textit{z}$\sim$3, as actually observed in BAT105 (KS p-value=0.74). We note that if we had instead assumed a constant X/R conversion given by the distribution observed at \textit{z}$\sim$1, the expected and observed distributions in BAT105 would have been inconsistent (KS p-value\textless10$^{-4}$).

%%%%%%%%%%%%%%%%%%%%%%%%%%%%%%%%%%%%%%%%%%%%%%%%%%%%%%%%%%%%%%%%%%%%%%%%%%%%%%%%%%%%%%%%%%%%%%%%%%

\section{Space densities: Radio vs X-ray}
\label{sec:density}

%Another difference that still has to be explained in high redshift blazars is the one related to their space density evolution. 
As already mentioned, X-ray and radio selected samples of blazars seem to find a different evolution, with radio selected blazars peaking at \textit{z}$\sim$2 (Mao17), similar to the RQ population, and X-ray selected ones at about \textit{z}$\sim$4 (\citealt{Ajello2009}, hereafter Aj09). We now want to update the Aj09 results, using the most recent version of the \textit{Swift}-BAT sample, and to compare them with the predictions based on the model described in the previous section.

\subsection{The blazar space density evolution from the 105-months BAT sample}

\begin{figure}
\centering
\includegraphics[width=\linewidth]{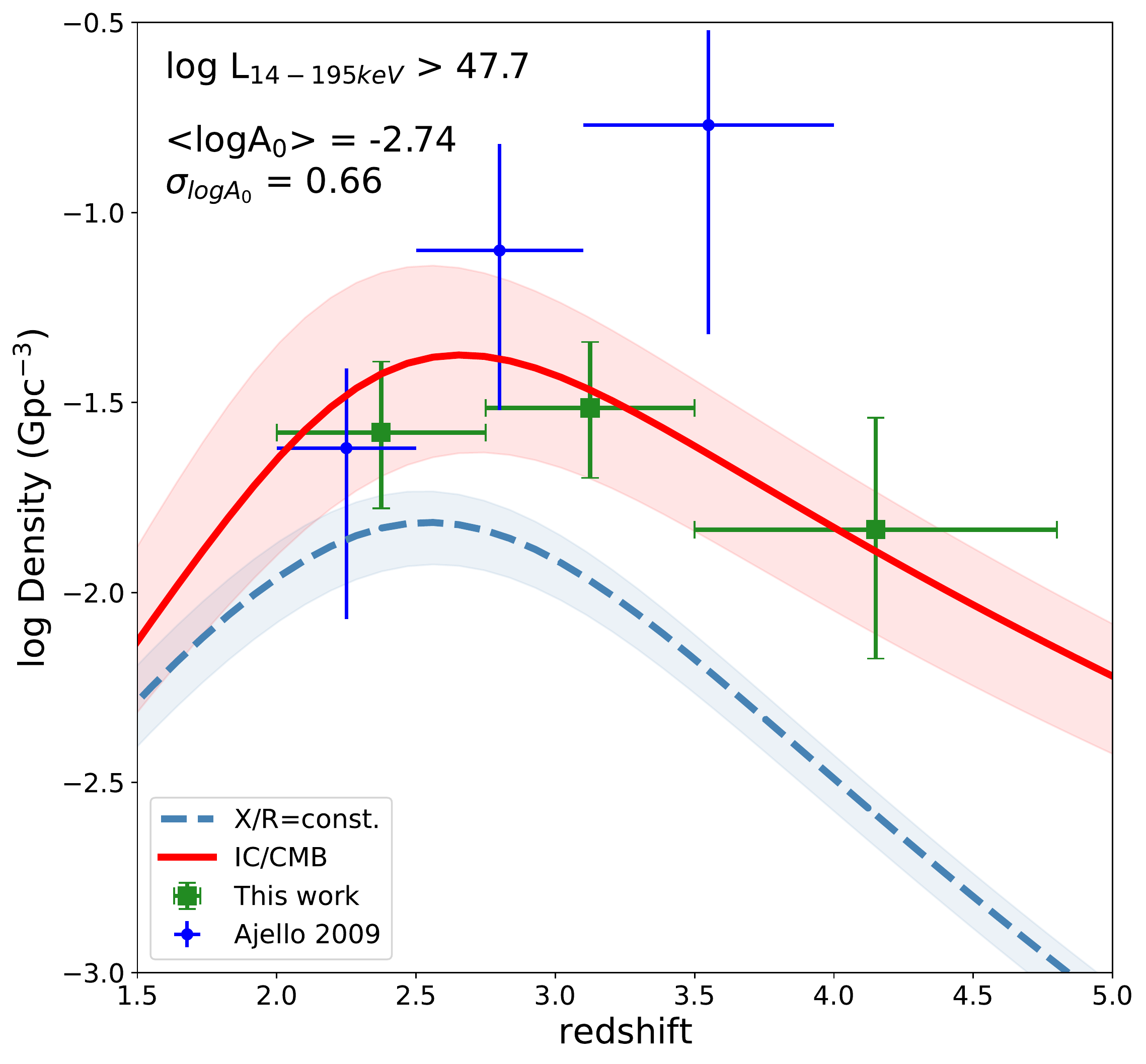}
\caption{Evolution of the X-ray space density of blazars as a function of redshift. The green squares are the space density obtained from the blazars in the 105-months BAT catalogue, while the space density computed by Aj09 using the 36-months BAT catalogue is reported with blue points. We also show the expected space density computed from the RLF discussed in Mao17 assuming both a constant X/R conversion factor (observed in the BZ1.5Jy, \textless X/R\textgreater = 1.95, $\sigma_\mathrm{X/R}=0.35$, dashed blue line) and a redshift evolving one which follows the trend predicted by the IC/CMB model derived in section \ref{sec:IC/CMBevo} (with parameters: \textless logA$_0$\textgreater=-2.74 and $\sigma_\mathrm{logA_0}$=0.66, solid red line). The shaded areas represent the variation of the expected space densities assuming a different photon index ($\Gamma=$1.5-1.6) and \textless logA$_0$\textgreater, $\sigma_\mathrm{logA_0}$ parameters (with a p-value\textgreater0.7).}
\label{fig:distr_all}
\end{figure}

The Aj09 work was based on an old version of the BAT catalogue (36 months), therefore, we decided to update the analysis using the most recent version that now includes data from the 105-months BAT catalogue \citep{Oh2018}.
Since the particular evolution has been observed in the most X-ray luminous blazars, logL$_\mathrm{{15-55keV}}$\textgreater 47.3 erg s$^{-1}$ (Aj09), we consider the same population of sources, that is, logL$_\mathrm{{14-195keV}}$\textgreater 47.7 erg s$^{-1}$ (where we assumed again $\Gamma$=1.54 for the conversion between the two bands) corresponding to 18 sources out of the 20 in the initial sample. Moreover, given the limiting flux of the BAT105 sample, we should be able to observe the majority of the sources within this luminosity range and with a photon index around 1.5 up to \textit{z}$\sim$4.5.
With this BAT105 sub-sample  we estimated the space density of blazars in the redshift range $z$=2--5  using the V$_\mathrm{{max}}$ method \citep{Schmidt1968}. After dividing the redshift range in three bins, each containing at least three sources, we computed the space density in each of them as follows:
\begin{equation}
    \rho (z_1, z_2) = \frac{1}{f_{sky}} \times \sum \frac{1}{V_{max}} 
    \label{eq:vmax}
\end{equation}
Where $f_\mathrm{sky}$ is the fraction of the sky here considered (=0.74) and the sum is on the objects within the redshift bin $z_1$--$z_2$, while V$_\mathrm{max}$ is the co-moving volume within which each source could have been observed. In other words, V$_\mathrm{max}$ is the co-moving volume between $z_1$ and the minimum between $z_2$ and the redshift at which the X-ray flux of the given source corresponds to the limit of the sample. The errors on each density estimate is then given by: $\frac{\sigma_\mathrm{P}}{N}$ $\times$ $\rho$($z_1$,$z_2$), where $N$ is the number of objects in the given redshift bin and $\sigma_\mathrm{P}$ the Poisson uncertainty on $N$.\\
In order to account for the unidentified  sources (i.e. without an associated redshift) in the BAT catalogue and the possibility that one or more of them are actually $z$\textgreater 2 blazars, we proceeded as follows. We applied to these objects the same criteria used to build the BAT105 sample, i.e. an X-ray flux above the completeness limit and a counterpart in the NVSS (see section \ref{sec:samples}). We then searched for possible identifications of the remaining sources in the recent literature, finding that the majority of them have been followed-up with dedicated spectroscopic observations within the \textit{BAT AGN Spectroscopic Survey} project (BASS; \citealt{Koss2017}) and turned out to be low-$z$ objects. Among the still unidentified sources ($\sim$20), only three of them have similar properties compared to the blazars here considered, namely a radio flux density $S_\nu$\textgreater 100 mJy. Since we cannot exclude that these three sources are in fact $z$\textgreater 2 blazars, we assumed them to have the same redshift distribution as the BAT105 objects with a similar radio flux. In particular, we computed the probability that one source has to belong to one of the different redshift bins here considered and then we increased the corresponding number of objects by this fraction times the number of possible blazars (3).
We report the values obtained in Tab. \ref{tab:sapce_den} and we show the space density evolution in Fig. \ref{fig:distr_all}  (green squares) together with the estimates of Aj09 (blue points) as a reference.
Interestingly, by sampling a larger co-moving volume and redshift range compared to Aj09 (up to \textit{z}$\sim$4.5, i.e. where the peak in Aj09 was expected), we found that the X-ray evolution is not as extreme as previously thought, with the space density peaking at redshift lower than 3.5 (the actual value of the peak is hard to constrain given the limited size of the sample). However, as explained in the next sub-section, we still find a discrepancy with respect to the space densities derived from radio surveys.

\begin{table}
\caption{X-ray space density of blazars with logL$_\mathrm{14-195keV}$\textgreater 47.7 erg s$^{-1}$ from the BAT105 sample. The values are also shown in Fig. \ref{fig:distr_all} (green squares).}
\centering
\setlength{\tabcolsep}{10pt}
\begin{tabular}{l cccc}
\hline
\hline
Redshift  bin & 2\textless$z$\textless2.75 &  2.75\textless$z$\textless3.5 & 3.5\textless$z$\textless4.8 \\
\hline
log $\rho$ (Gpc$^{-3}$) &  -1.58$_{-0.20}^{+0.19}$ & -1.51$^{+0.17}_{-0.18}$ & -1.83$^{+0.29}_{-0.34}$\\
\hline
Number of sources  & 7 & 8 & 3 \\
\hline
\hline
\end{tabular}
\label{tab:sapce_den}
\end{table}

\subsection{Comparing X-ray and radio results}

In order to compare the X-ray space density computed few lines above to the one derived from radio observations, we considered once again the RLF determined by Mao17. Assuming a constant X/R conversion factor such as the one observed in the BZ1.5Jy sample ($\bar{z}\sim1.1$), we converted the RLF to an X-ray LF (XLF), which can be expressed in terms of the radio one as follows:
\begin{equation}
    \Phi_X(L_X,z) \times L_X = \Phi_R(L_R,z) \times L_R 
    \label{eq:lum_fun_x}
\end{equation}
Where, in this case, $L_\mathrm{X}$ indicates the luminosity in the 14-195~keV energy band. Since the X-ray luminosities used to compute the X/R ratios for the C19 and the BZ1.5Jy samples are measured in the 2-10~keV energy range, we assumed again a conversion factor associated to a photon index $\Gamma=1.54$.
As shown in Fig. \ref{fig:distr_all} (dashed blue line), in this case the expected number of blazars is significant smaller compared to the estimates from the BAT105 sample, with the space density differing by a factor of $\sim$2 (6) at \textit{z}$\sim$2.3 (4.2).\\
In order to verify if the interaction with the CMB could reconcile the two space density evolutions, when computing the XLF in eq. \ref{eq:lum_fun_x} we also considered radio-to-X-ray conversion factors given by eq. \ref{eq:enhace} where the parameter A$_0$ follows the best-fit distribution obtained in the previous section. We report in Fig. \ref{fig:distr_all} the X-ray space density obtained from the integration of the XLF considering the effect of CMB (solid red line). 
The shaded areas show the expected evolutions assuming a different photon index ($\Gamma$ in the range 1.5-1.6, the median values observed at energies 20-100~keV in a radio selected sample of blazars, \citealt{Langejahn2020}) and a different distribution of the logA$_0$ parameter which can still well reproduce the observed C19 X/R ratios (p-value\textgreater0.7, see section \ref{sec:IC/CMBevo}).
As it is clear from Fig. \ref{fig:distr_all}, the effect of the CMB derived from the properties of \textit{z}\textgreater4 radio selected blazars seems to be able to reconcile the cosmological evolutions observed in the radio and in the X-ray bands by increasing the expected space density of X-ray selected blazars, especially at high redshift. 
It is worth noting that the IC/CMB model does not predict a significant shift of the space density peak computed from X-ray surveys, but only a slower decrease after the peak moving towards higher redshift due to the additional X-ray emission coming from the interaction with the CMB photons. As shown above, this is in agreement with our new analysis of the latest BAT data-set. Larger X-ray selected samples able to better constrain the peak, as described in the next section, will be instrumental to test this critical prediction.

%%%%%%%%%%%%%%%%%%%%%%%%%%%%%%%%%%%%%%%%%%%%%%%%%%%%%%%%%%%%%%%%%%%%%%%%%%%%%%%%%%%%%%%%%%%%%%%%%%
\section{Predictions for the \lowercase{e}ROSITA mission}
\label{sec:predictions}

\begin{figure*}
\centering
\includegraphics[width=0.49\linewidth]{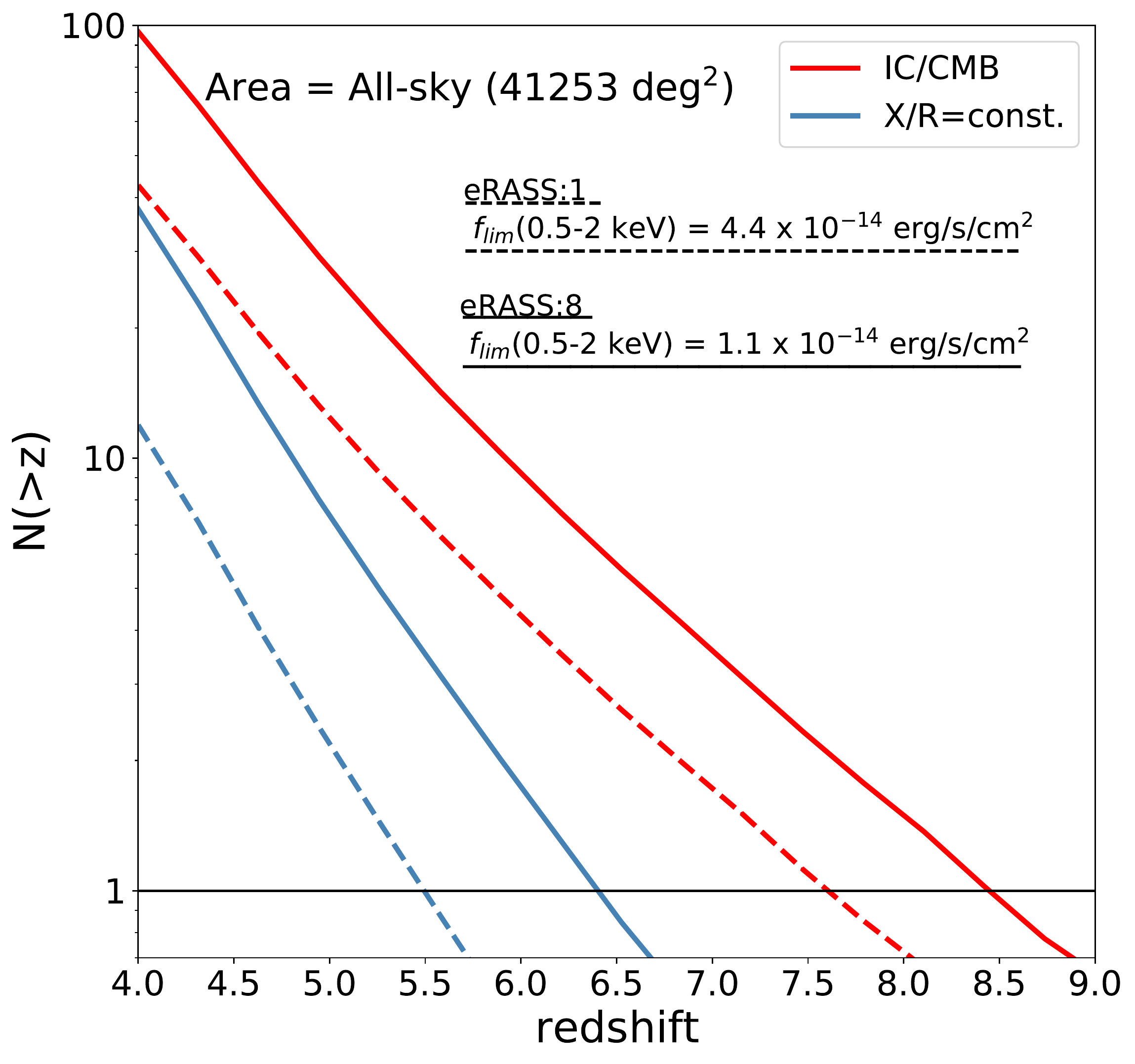}
\includegraphics[width=0.485\linewidth]{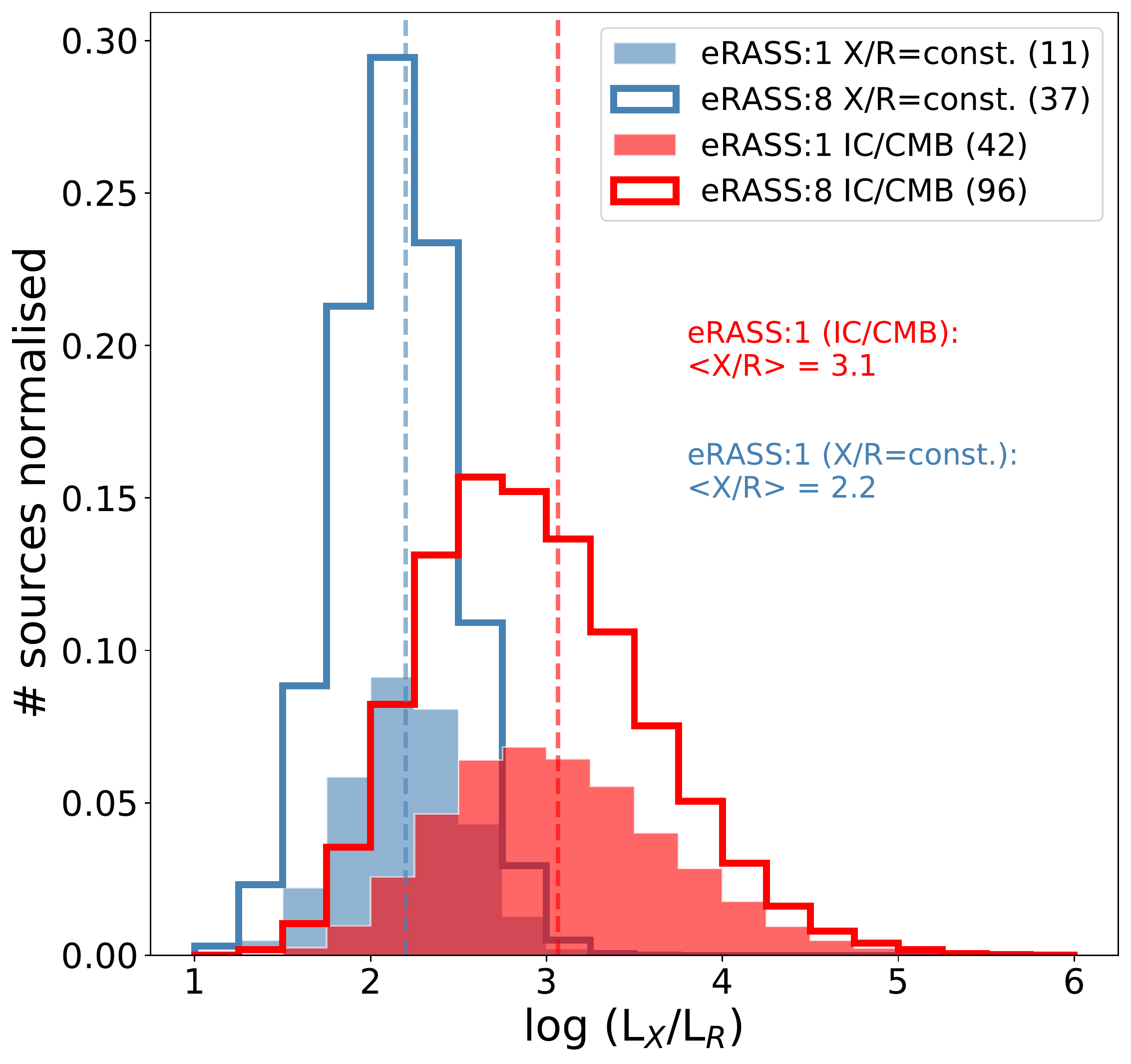}
\caption{Prediction of the number (left panel) and properties (right panel) of $z$\textgreater4 blazars expected to be found with the eROSITA all-sky survey. In both panels red lines represent the estimates made considering the IC/CMB model, whereas blue lines indicate the estimates assuming non-evolving X/R ratios. Dashed lines and filled histograms refer to the predictions after the first scan of the sky (eRASS:1), while the continuous lines and empty histograms are for 8 scans (eRASS:8). \textbf{Left:} Expected number of detected blazars above a given redshift as a function of redshift. \textbf{Right:} X-ray-to-radio ratio distributions. The dashed vertical lines indicate the mean values expected in the eRASS:1. Both filled and empty histograms have been normalised dividing by the total number of blazars expected in the eRASS:8 considering the corresponding evolving scenario. In the legend, the total number of blazars expected to be observed for each scenario is reported in brackets .}
\label{fig:predictions}
\end{figure*}

An important observational consequence of the IC/CMB model is the presence of many X-ray powerful blazars at high redshifts. Therefore one way to further test this scenario is to focus on very high redshift (z\textgreater4) blazars, where the effect related to the CMB is expected to be dominant.
To date the most suitable X-ray survey to build a statistically relevant sample of high-z blazars is the \textit{extended ROentgen Survey with an Imaging Telescope Array} (eROSITA; \citealt{Merloni2012}). 
Having the maximal sky coverage possible and a relatively high  sensitivity (from 4.4 $\times \, 10^{-14}$ erg/s/cm$^2$ after the first scan, e-RASS:1, to 1.1 $\times \, 10^{-14}$ erg/s/cm$^2$ after 8 scans, eRASS:8, in the 0.5-2~keV energy band) this mission is ideal for the detection of a large number of X-ray powerful sources at high-$z$ (e.g. \citealt{Medvedev2020,Wolf2021, Khorunzhev2021}).
Following the approach outlined in section \ref{sec:IC/CMBevo} for the BAT105 sample, we computed the number and properties of the blazars expected to be found by the eROSITA mission. As shown in Fig. \ref{fig:predictions} (left side), the potential number of observed blazars at $z$\textgreater4 is significantly different in an IC/CMB scenario (with the parameters discussed in the previous sections) or in the case the X/R ratios do not evolve with redshift. In particular, if the IC/CMB interaction plays a significant role in the X/R ratios evolution with redshift, we expect to detect about three times more blazars above the eROSITA flux limit compared to a non-evolving scenario ($\sim$100 versus $\sim$40 sources in the final e-RASS:8). Another important difference  between the two scenarios is that, according to an IC/CMB evolution, we are also expecting to find more blazars in the so-called epoch of re-ionisation (EoR; \textit{z}$\gtrsim$6) where only one blazar has been discovered thus far \citep{Belladitta2020}.
Moreover, also the X-ray and radio properties of the detected blazars will be instrumental to check whether their X/R ratios evolve according to the IC/CMB scenario or not. As shown in Fig. \ref{fig:predictions} (right panel), we expect to observe significantly different values of X/R ratios in the IC/CMB case, with a large number of sources with very extreme X/R values (\textgreater3). Detailed studies on these extreme sources, such as the ones performed on the extended jets of X-ray-bright sources at intermediate redshifts (\textit{z}$\sim$3, e.g. \citealt{Worrall2020}) with the \textit{Chandra} telescope, will be instrumental to further constrain the IC/CMB mechanism.

%%%%%%%%%%%%%%%%%%%%%%%%%%%%%%%%%%%%%%%%%%%%%%%%%%%%%%%%%%%%%%%%%%%%%%%%%%%%%%%%%%%%%%%%%%%%

\section{Conclusions}
\label{sec:conclusions}

In this work we tested whether the recently observed increase of the X/R ratios in high-$z$ blazars \citep{Ighina2019} can be attributed to a \textit{fractional} interaction between the relativistic electrons in the jet with the CMB photons. We found that, by assuming an intrinsic scatter around the average expected trend (meant to represent the different physical properties of the blazar population) we can reproduce the X/R distributions observed at different redshifts, including the extreme values of the $z$\textgreater2 blazars detected by the \textit{Swift}-BAT X-ray telescope.
Moreover, after estimating the X-ray space density evolution of blazars from the latest version of the BAT catalogue (105-months) we found that the same IC/CMB model could also explain the differences between X-ray and radio evolutions of the blazar population reported in the literature \citep{Ajello2009,Mao2017}. 
Our results show that the different evolution observed in the radio and in the X-rays can be attributed to the variation of the X-ray-to-radio luminosity ratio expected from the IC/CMB interaction. However, we cannot exclude that this variation may be related to a different effect, like a significant change in the host galaxy properties with $z$ for instance \citep{Wu2013}, or even a combination of different effects.\\
Another crucial way to test and constrain the importance of the IC/CMB interaction will be to focus on the numerous sources at high redshift expected to be found in wide-area X-ray surveys such as those carried out with the eROSITA mission. 
High angular resolution observations on the most extreme sources with the \textit{Chandra} telescope will be another important and complementary way to constrain the IC/CMB interaction (e.g. \citealt{Simionescu2016,Worrall2020}). 
The systematic study of extended X-ray emission in these high-$z$ blazars will represent the final confirmation of the relevance of the IC/CMB mechanism on the X-ray emission of high redshift AGNs.

\section*{Acknowledgements}
We want to thank Massimo Dotti for his help in the last couple of years. We also want to thank Gabriele Ghisellini and Lea Marcotulli for helpful discussions and the referee for their useful comments. We acknowledge financial contribution from the agreement ASI-INAF n. I/037/12/0 and n.2017-14-H.0 and from INAF under PRIN SKA/CTA FORECaST. This work made use of data supplied by the UK \textit{Swift} Science Data Centre at the University of Leicester. This research made use of Astropy, a community-developed core Python package for Astronomy \citep{astropy2018}. %This research has made use of the SIMBAD database, operated at CDS, Strasbourg, France \citep{Simbad2000}.

%%%%%%%%%%%%%%%%%%%%%%%%%%%%%%%%%%%%%%%%%%%%%%%%%%%%%%%%%%%%%%%%%%%%%%%%%%%%%%%%%%%%%%%%%%%%
\section*{Data Availability}

The majority of the data used in the paper are publicly available as described in the main text. Reprocessed data are available upon request to the main author.



\bibliographystyle{mnras}
\bibliography{biblio} 

%%%%%%%%%%%%%%%%%%%%%%%%%%%%%%%%%%%%%%%%%%%%%%%%%%

%%%%%%%%%%%%%%%%% APPENDICES %%%%%%%%%%%%%%%%%%%%%

\appendix

\section{GB6J1711+3830 X-ray analysis}
\label{app:GB6J1711+38}
\begin{figure}
\centering
\includegraphics[width=\linewidth]{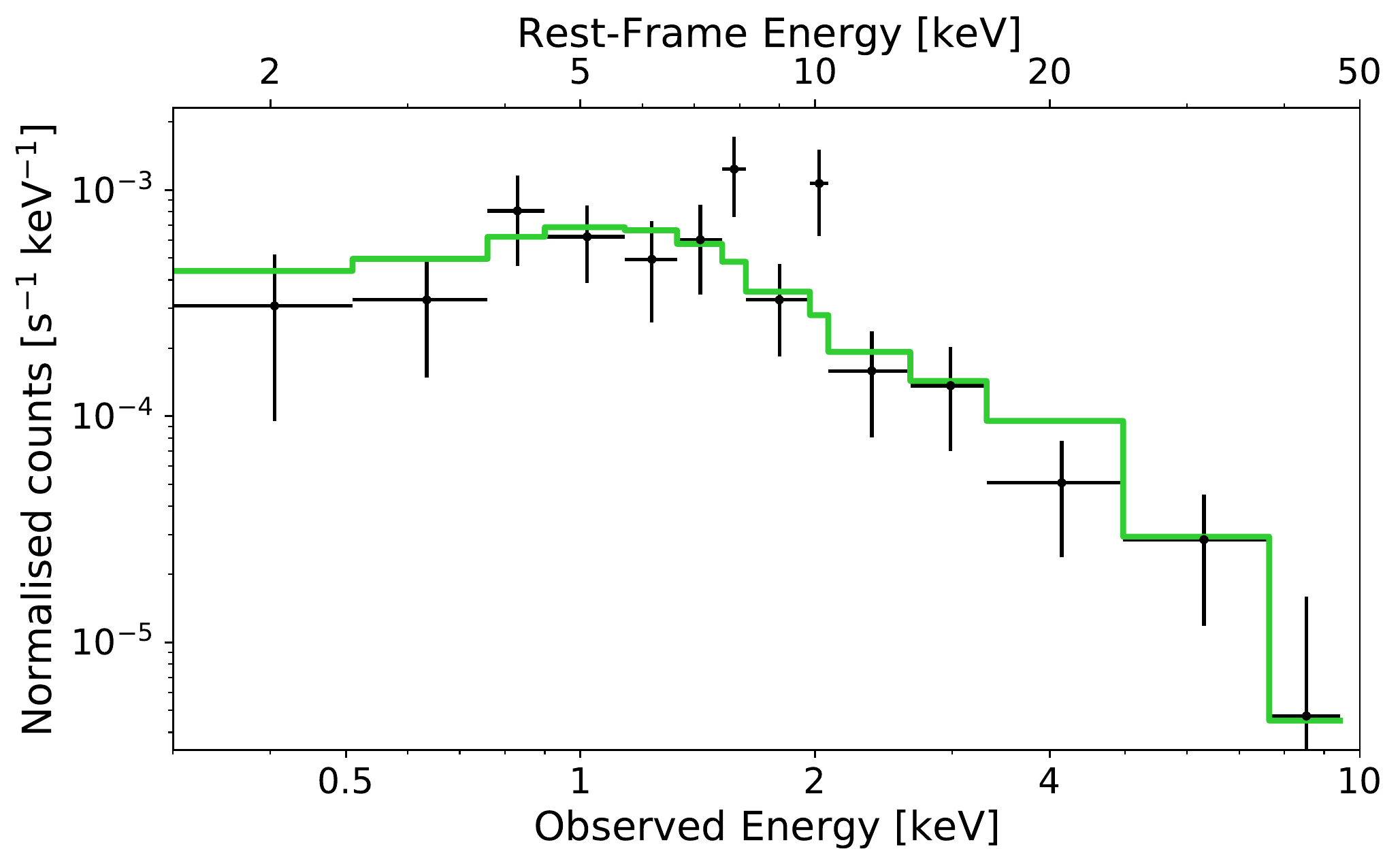}
\includegraphics[width=\linewidth]{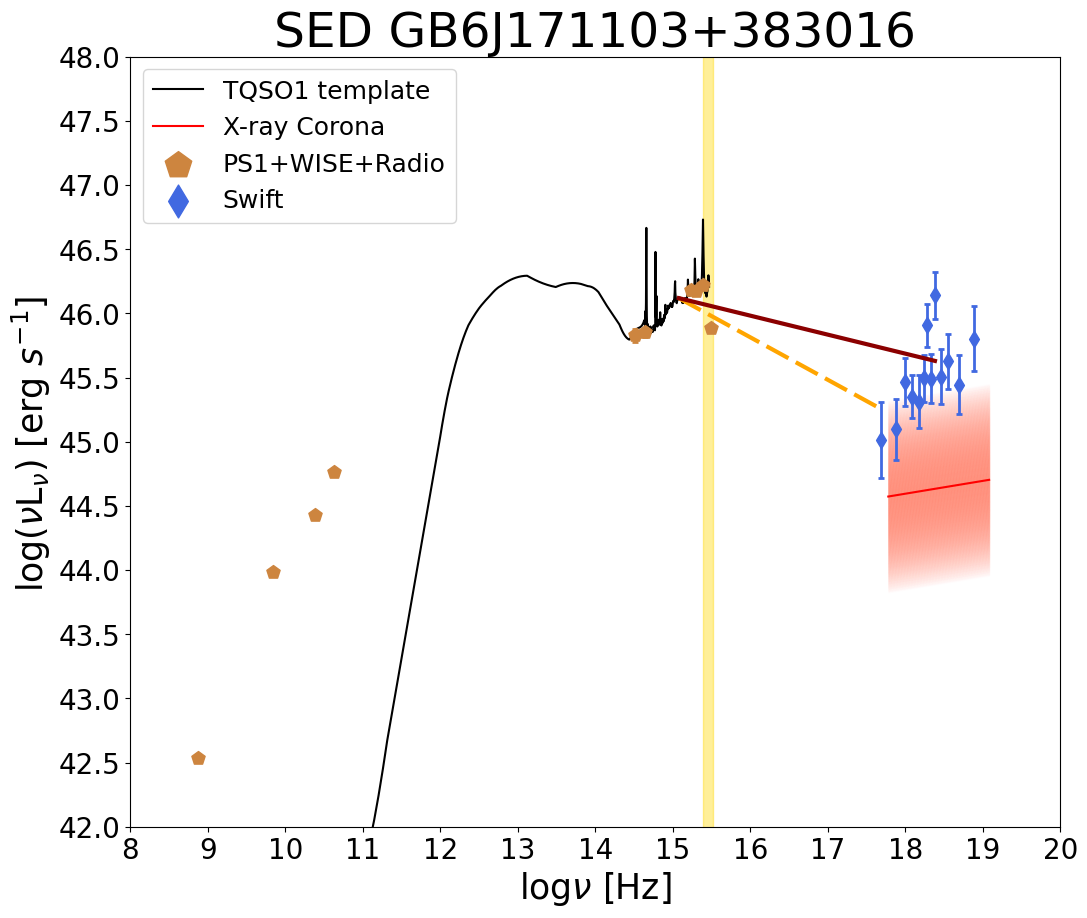}

\caption{\textbf{Top:} X-ray spectrum of GB6J1711+3830 from the \textit{Swift}-XRT dedicated observation modelled with a power law with only Galactic absorption (solid green line). The data have been binned at 3$\sigma$ significance for graphical purposes only. \textbf{Bottom:} Broad-band spectral energy distribution of GB6J1711+3830. The X-ray data from \textit{Swift}-XRT are reported as blue diamonds, while the radio and optical data (brown pentagons) are described in the text. The black line is a template for the optical emission of quasars \citep{Polletta2007} while the red line and the corresponding shaded region indicate the expected X-ray emission from the X-ray corona in a RQ AGN \citep{Steffen2006}.}
\label{fig:GB6J1711+38}
\end{figure}

In this section we report the X-ray analysis of the source: GB6J1711+3830. This is a $z$=4.0 AGN (the optical spectrum is reported in C19) classified as a blazar candidate on the basis of its radio properties. This is the only object from \cite{Caccianiga2019} whose X-ray analysis was not included in \cite{Ighina2019} because it had not been observed at the time of publication; we discuss the analysis and its X-ray properties here. The dedicated 50 kilo-seconds X-ray observation was carried out between October and November 2020 with the \textit{Swift}-XRT telescope (Obs. ID: 3110834, P.I. Belladitta). We analysed it following the same procedure used for the rest of the C19 sample, that is, using the \texttt{XSPEC} (v12.11.1) package and considering a simple power law absorbed by the Galactic column density along the line of sight (N$_H$=4.07$\times$10$^{20}$ cm$^{-2}$; \citealt{HI4PI2016}). Combining the results of this analysis with optical information from the Panoramic Survey Telescope and Rapid Response System (m$_i$=20.46$\pm$0.01; Pan-STARRS, \citealt{Chambers2016}) we also computed the rest-frame ratio between the X-ray (10~keV) and the UV (2500~\AA) luminosities, the $\tilde{\alpha}_{ox}$ parameter\footnote{$\tilde{\alpha}_{ox}$=-0.303 log$\frac{L_{10keV}}{L_{2500\AA}}$} \citep{Ighina2019}. Adopting the classification used for the rest of the C19 sample based on the shape ($\Gamma$\textless1.8) and the relative X-ray luminosity compared to the optical one ($\tilde{\alpha}_{ox}$\textless1.355) we consider GB6J1711+3830 as a bona-fide blazar in this work. We report in Tab. \ref{tab:GB6J1711+38} the results of the analysis and in Fig. \ref{fig:GB6J1711+38} the observed X-ray spectrum (top panel) and its rest-frame multi-wavelength SED (bottom panel). Optical data are from PanSTARRS and radio data from the Giant Metrewave Radio Telescope all-sky survey (150~MHz; \citealt{Intema2017}), NVSS (1.4~GHz), the Green-Bank Survey (5~GHz; \citealt{gregory1996}) and the The Karl G. Jansky Very Large Array (8.4 GHz). The shaded red region represents the expected X-ray luminosity of a RQ AGN considering the L$_{UV}$--L$_X$ relation, and its 1$\sigma$ uncertainty, reported in \cite{Steffen2006}, whereas the yellow column indicates the region with a significant dropout of the luminosity caused by the Lyman absorption (912--1216 \AA). The slopes of the solid red and dashed orange lines are given by 1-$\tilde{\alpha}_{ox}$ and 1-${\alpha}_{ox}$\footnote{${\alpha}_{ox}$=-0.384 log$\frac{L_{2keV}}{L_{2500\AA}}$} respectively. The optical-IR template in black is taken from the SWIRE template library \citep{Polletta2007}.

\begin{table}
\caption{Results of the X-ray analysis on GB6J1711+3830.}
\begin{threeparttable}
\centering
\begin{tabular}{ccccccc}
\hline
\hline
$\Gamma$    &   f$_\mathrm{0.5-10keV}^\mathrm{a}$ &   L$_\mathrm{2-10keV}^\mathrm{b}$   &   $\tilde{\alpha}_{ox}$  & cstat / d.o.f.\\
\hline
1.46$^{+0.31}_{-0.31}$  & 	8.26$_{-1.62}^{+2.06}$  &   3.84$_{-0.57}^{+0.52}$  & 1.148$_{-0.043}^{+0.053}$ & 77.7 / 66 \\
\hline
\hline
\end{tabular}
\label{tab:GB6J1711+38}

\begin{tablenotes}
\item Errors are reported at 90\% level of confidence.
\item [a] in units of 10$^{-14}$ erg s$^{-1}$ cm$^{2}$
\item [b] in units of 10$^{45}$ erg s$^{-1}$
\end{tablenotes}

\end{threeparttable}
\end{table}

%%%%%%%%%%%%%%%%%%%%%%%%%%%%%%%%%%%%%%%%%%%%%%%%%%

% Don't change these lines
\bsp	% typesetting comment
\label{lastpage}
\end{document}